# Defect Management Strategies in Software Development

Suma V and Gopalakrishnan Nair T.R.
*Research and Industry Incubation Centre, Dayananda Sagar Institutions
Bangalore, India*

## 1. Introduction

Software is a unique entity that has laid a strong impact on all other fields either related or not related to software. These include medical, scientific, business, educational, defence, transport, telecommunication to name a few. State-of-the-art professional domain activities demands the development of high quality software. High quality software attributes to a defect-free product, which is competent of producing predictable results and remains deliverable within time and cost constraints. It should be manageable with minimum interferences. It should also be maintainable, dependable, understandable and efficient. Thus, a systematic approach towards high quality software development is required due to increased competitiveness in today's business world, technological advances, hardware complexity and frequently changing business requirements.

### 1.1 Software Engineering

Software Engineering is a discipline that aims at producing high quality software through systematic, well-disciplined approach of software development. It involves methods, tools, best practices and standards to achieve its objective. The three main phases of software development life cycle (SDLC) are requirement analysis, design and implementation phase (Roger S. Pressman, 2005, Ian Somerville, 2006). To deploy high quality software, it is essential to develop a defect-free deliverable at each phase.

A defect is any blemish, imperfection, or undesired behaviour that occurs either in the deliverable or in the product. Anything related to defect is a continual process and not a state.

### 1.2 Need for defect management

Defect analysis at early stages of software development reduces the time, cost and resources required for rework. Early defect detection prevents defect migration from requirement phase to design and from design phase into implementation phase (Jeff Tian, 2005). It enhances quality by adding value to the most important attributes of software like reliability, maintainability, efficiency and portability. Hence, industry should go for defect management at every stage of development to gain total confidence with customers (Watts

S. Humphrey, 1989; Kashif Adeel et al., 2005; Vasudevan S, 2005; Mukesh Soni, 2006; Purushotham Narayan, 2003). The two approaches of defect management are i) defect detection and ii) defect prevention (DP). Defect detection techniques identify defect and its origin. Defect prevention is a process of minimizing defects and preventing them from re-occurrence in future.

**1.3 Conventional defect detection strategies**

There are several approaches to identify defects like inspections, prototypes, testing and correctness proof. Inspection is examination of human artefacts to detect defects at the early stages of software development. It is the most effective and efficient quality assurance technique. A prototype is an experimental version of software release. It helps both customer and developer to verify that the product meets all the stipulated requirements. It enables both parties to resolve ambiguous requirements to a well-defined specification. Thus, prototyping eliminates defects caused due to ambiguity. Testing is quality control activity that identifies defects at the time of implementation. It uncovers those defects, which could have escaped by identification at the early stages of development. Correctness proof discovers defects at coding stage. The code that fails to meet the requirements of correctness proof indicates existence of defect (Vasudevan. S, 2005).

**1.4 Defect classification**

Defect classification follows defect detection activity. Two occasions in which defects usually classified are i) defect injection time and ii) defect fixing time. Several models and tools assist in defect classification. Orthogonal Defect Classification (ODC) is the most popular technique that groups defects into types rather than considering them as individuals. It helps to identify process areas that require attention (Chillarege et al., 1992). Another popular approach for defect classification is HP model of defect origins, types and modes. This model links together defect types and origin by identifying type of defect appearing at the origin (Stefan Wagner, 2008). Yet, another technique of defect classification considers certain factors like logical functions, user interface, standards and maintainability and so on. Further, each company has its own methodology of classifying defects.
Identified defects can be categorized depending on defect type. They are blocker type of defects which prevent continued functioning of the developer team, critical type that results in software crash, system hang, loss of data etc. Defect is categorized as a major type when a major feature collapses and a minor type when defect causes a minor loss of function, still allowing an easy work around. Trivial category of defect arises due to cosmetic problems. Based on these categories, severity levels are assigned as either urgent/show stopper, medium/work around or low/cosmetic (Vasudevan S, 2005).

**1.5 Defect distribution**

Depending on the sequence of process in which software can be dealt with different phases of SDLC, defects can be accounted based on the phases in which they occur. An empirical study conducted across several projects from various service-based and product-based organizations reveals that requirement phase contains 50% to 60% of total defects. 15% to 30% of defects are at design phase. Implementation phase contains 10% to 20% of defects. Remaining are miscellaneous defects that occurs because of bad fixes. Bad fixes are injection of secondary defects due to bad repair of defects. Table 1. is a sampled data obtained from several leading software industries. This depicts time and defect profile at each phase of development.

|  | P1 | P2 | P3 | P4 | P5 | P6 |
|---|---|---|---|---|---|---|
| Total project time (in man hours) | 250 | 507 | 2110 | 4786 | 6944 | 9220 |
| Total requirement time | 25 | 55 | 800 | 2047 | 2597 | 2550.6 |
| Total inspection time | 2 | 6 | 48 | 163 | 208 | 204 |
| Total testing time | 5 | 16 | 80 | 575 | 281 | 821 |
| Total number of defects | 30 | 77 | 139 | 200 | 254 | 375 |
| Number of defects identified by inspection | 16 | 40 | 68 | 123 | 112 | 225 |
| Number of blocker type of defect | 3 | 8 | 15 | 19 | 25 | 40 |
| Number of critical type of defect | 4 | 9 | 14 | 24 | 30 | 42 |
| Number of major type of defect | 6 | 15 | 30 | 40 | 51 | 71 |
| Number of minor type of defect | 7 | 17 | 34 | 47 | 48 | 75 |
| Number of trivial type of defect | 10 | 28 | 46 | 70 | 100 | 147 |
| % of defects at requirements phase | 51.72 | 59.23 | 57.92 | 59.17 | 56.32 | 56.39 |
| Total amount of design time | 46 | 110 | 400 | 1323 | 1966 | 3080 |
| Total amount of inspection time | 5 | 11 | 48 | 105.84 | 157 | 246 |
| Total amount of testing time | 11 | 21 | 112 | 158.76 | 236 | 369 |
| Total number of defects | 10 | 26 | 55 | 75 | 120 | 182 |
| Number of defects identified by inspection | 5 | 14 | 24 | 33 | 77 | 102 |
| Number of blocker type of defect | 1 | 2 | 6 | 7 | 12 | 16 |
| Number of critical type of defect | 1 | 6 | 7 | 15 | 18 | 32 |
| Number of major type of defect | 2 | 5 | 12 | 15 | 25 | 44 |
| Number of minor type of defect | 2 | 5 | 11 | 16 | 29 | 42 |
| Number of trivial type of defect | 4 | 8 | 19 | 22 | 36 | 48 |
| % of defects at design phase | 17.24 | 20.00 | 22.92 | 22.19 | 26.61 | 27.37 |
| Total amount of implementation time | 101 | 165 | 640 | 756 | 1300 | 2200 |
| Total amount of inspection time | 10 | 23 | 112 | 90.72 | 156 | 264 |
| Total amount of testing time | 36 | 45 | 200 | 113.4 | 195 | 330 |
| Total number of defects | 8 | 17 | 36 | 53 | 67 | 98 |
| Number of defects identified by inspection | 4 | 9 | 24 | 27 | 37 | 54 |
| Number of blocker type of defect | 1 | 1 | 3 | 3 | 5 | 10 |
| Number of critical type of defect | 2 | 4 | 4 | 12 | 14 | 21 |
| Number of major type of defect | 2 | 4 | 9 | 13 | 13 | 22 |
| Number of minor type of defect | 1 | 3 | 8 | 12 | 16 | 22 |
| Number of trivial type of defect | 2 | 5 | 12 | 13 | 19 | 23 |
| % of defects at implementation phase | 13.79 | 13.08 | 15.00 | 15.68 | 14.86 | 14.74 |

Table 1. Sampled data obtained from leading software industries        P = Project

**1.6 Defect pattern of various types of defects**
Table 1. indicates the existence of various types of defect pattern at each phase of software development. Table 2. indicates percentage of possibility of occurrences of various defect patterns.

| Type of defect | Possibility % of defect occurrences | Overall defect pattern |
| --- | --- | --- |
| Blocker type | 5% to 15% | 10% |
| Critical type | 10% to 25% | 20% |
| Major type | 20% to 25% | 25% |
| Minor type | 10% to 20% | 15% |
| Trivial type | 15% to 55% | 30% |

Table 2. Defect pattern in software development

**Rationale for defect occurrences**
It is a universally accepted fact that nothing can be created with absolute perfection. Software engineering is not an exception. Thus, there inevitably exists rationale for defect occurrences. The common causes for defect occurrences at requirements phase are requirement incompleteness, inconsistency, ambiguity, requirement change and requirement presentation. The common reasons for defect occurrences at design phase are non-conformance to external specifications, internal specifications, logical specifications, interface specifications, component specification, security, organizational policies and standards in addition to non–conformance to design with requirement specification. The common sources for defect occurrences at implementation phase are improper error handling, improper algorithm, programming language shortcomings, wrong data access and novice developers (Purushotham Narayan, 2003).

**1.7 Root cause analysis for defect pattern**
Root Cause Analysis (RCA) is an effective technique to investigate the origin for defect injection (David N. Card, 2006). This analysis helps to prevent reoccurrences of defect in future. Three general classifications of tools that support root cause analysis are i) problem solving tools ii) management and planning tools iii) product development and process improvement tools. Problem solving tools includes pareto charts, check sheet, cause-and-effect-diagram (CED), histograms, scatter plots, trend analysis graphs and control charts. They are basic tools of quality. They break potential root cause into more detailed root causes to identify all related factors for defect occurrence. Management and planning tools includes affinity diagrams, relations diagrams, matrix data analysis charts, hierarchy diagrams, matrices and tables to display value and priority, precedence diagrams. They establish quantifying interrelationship between potential root causes and factors driving these root causes. Product development and process improvement tools include failure modes and effect analysis (FMEA), fault tree analysis (FTA) and potential problem analysis and current reality tree (CRT) (Anthony Mark Doggett, 2004).
Root cause analysis is either logical analysis or statistical analysis. Logical analysis establishes a logical relation between effect and cause. This is usually human intensive analysis that demands expertise knowledge of product, process, development and environment. Statistical analysis establishes the probability relation of effect and cause. This relation depends upon empirical studies of similar projects or from evidences collected within the project (Jeff Tian, 2001).

### 1.8 The most common root cause classification

Despite the existence of various rationales, RCA techniques enable to classify the most common root causes and percentage of their contributions towards various defect patterns. They are communication (25% to 30%), education (20% to 25%), oversight (30% to 40%), transcription (20% to 25%) and miscellaneous (5% to 10%). From the defect distribution and defect pattern analysis, it is evident that trivial defects contribute more towards defect injection.

## 2. Defect prevention

Awareness of defect injecting methods and processes enables defect prevention. It is the most significant activity in software development. It identifies defects along with their root causes and prevents their reoccurrences in future.

### 2.1 Benefits of defect prevention

Defect prevention is vital for the successful operation of the industry. Since three decades, the benefits of DP are widely recognised and remedial measures are taken to overcome the impact of defect on quality. Inspection is cost effective as defects get uncovered in the early developmental phases, adds value to the dependency attributes of software like maintainability, availability and reliability. It enhances quality and gains customer satisfaction at all levels. It's adherence to meet the committed schedules further enhances the total productivity. It reflects the maturity level of the company and builds up the team spirit (Caper Jones, 2008). It is process, product and team appraisal activity and a mechanism for propagating the knowledge of lessons learned between projects or between various phases of software development (Van Moll, 2002).

Therefore, it is imperative to introduce DP at every stage of software life cycle to block defects at the earliest. It is necessary to take corrective actions for its removal and avoidance of its reoccurrence.

### 2.2 Conventional defect prevention strategies

Since the inception of DP activities in industry, several strategies have evolved towards their implementation. They are coined as conventional DP strategies.
The three conventional defect prevention strategies are
      i. Product approach of defect prevention
      ii. Process approach of defect prevention
      iii. Automation of development process  (Jeff Tian ,2001)

i) The three product approaches of defect prevention techniques are defect prevention through error removal technique, defect reduction through fault detection and removal technique and defect containment through failure prevention technique.

**DP through error removal technique**
Defects occurring due to human actions are removed by following any of the following techniques

384                                                                    Recent Advances in Technologies*Train and educate the developers*
Nearly 50% to 75% of the defects are due to human actions. Therefore, development team need training and education in product and domain specific knowledge. Further, an effective DP emphasize on a systematic approach of system development. Introduction of best practices like clean room approach, personal software process and team software reduces defect injection.

*Use of formal methods like formal specification and formal verification*
Formal methods consists of formal specification and formal verification techniques for defect detection. Formal specification uses formal logic and discrete mathematics to check for ambiguous, inconsistent and incomplete requirement specification. Formal verification verifies design constructs against the validated requirement specification. This avoids injection of accidental defects.

*DP based on tools, technologies, process and standards*
Defect injection reduces with use of object-oriented technology, follow up of well-defined process, right choice of tools and adherence to appropriate standards for product development

*Prevention of defects by analyzing the root causes for defects*
Root cause analysis is the most effective method of addressing defect. It is a periodic assessment to identify the root causes of defects with the aid of tools and methods like cause/effect diagrams, pareto analysis etc. Implementations of corresponding corrective actions along with preventive actions eradicate future defects.

**Defect reduction through fault detection and removal technique**
Organizations that develop safety critical projects and complex projects adopt fault detection and removal technique. Inspection is a static technique of fault detection and removal that examines human artefacts to detect and eliminate static defects. This prevents defect migration into later phases of development and consequently its manifestation. Testing is a dynamic activity that detects and eliminates dynamic defects that occur in software product during the development process. Testing includes all tests from unit test up to beta test.

**Defect containment through failure prevention technique**
Defect containment is a technique either to eliminate the causal relation that exits between fault and failure or to minimize the impact of their relation. As a result, faults continue to reside in the product but prevent defects. Techniques used for this purpose include recovery blocks, n-version programming, safety assurance and failure containment.

ii) From the perception of process approach of defect prevention technique, the management of software industry holds certain responsibilities in DP. Some of the actions that are handled as described in process change management key process area are – goals, commitment to perform, ability to perform, activities performed, measurements and analysis and verifying implementations (Pankaj Jalote, 2002).

*Goals*
The organization establishes goals like plan for DP, identify common causes for defects, prioritize the common causes and take corrective actions to eliminate them.

*Commitment to perform*
Implementation of these goals appears in the form of written policies both for organization and for the product. It includes long-term plans for financial, human and any other resource support that are required for DP activities. Implementation of the DP activities and continual review of them forms a part of organizational policy.

www.intechopen.com

*Ability to perform*
In accordance with the Key Process Area, an organizational level team and project level team exists to perform DP activities. A schedule to perform DP activities is prepared. The plan describes the task kick-off meetings, causal analysis meetings, implementing actions and their review, management participation, training activities, tools suitable to perform DP activities etc.

*Activities performed*
Performance of DP activities complies with the scheduled plan. DP activities include recommendation of corrective actions for defects, documentation, review and verification of the DP action items, control and management of DP data.

*Measurements and analysis*
Measurements confirm the status of DP activities. Knowledge of fundamentals of measurement and analysis forms the key path to success. Measurement and analysis enables the management to gain process insight of their organization.

*Verifying implementations*
Verification of implementation includes review of DP activities on a periodic basis with management, project managers and quality assurance group. It helps to accomplish continual process improvement in the organization.

iii) DP through automation of development process
There are many tools in usage for managing defects. Currently, several methods are under development that can detect and manage defects in an autonomic nature. Automation eliminates human intensive defects. Therefore, automation of development process is another approach towards DP. Automation tools are available from requirements phase to testing phase. Tools for automation purpose at requirements phase are quite expensive. Automation of attributes like consistency check is possible while attributes like completeness check may not be possible to automate completely. Tools used at this phase include requirement management tool, requirements recorder tool, requirement verifier's tool etc. Design tools include database design tool, applications design tool, visual modelling tool like Rational Rose etc. Automation of testing phase is by the use of tools like code generation tool, code-testing tool and code-coverage-analyzer tool. Several tools like defect tracking tool, configuration management tool and test procedures generation tool are functional at all phases of development (Elfriede Dustin at el., 1999; 2009).

**2.3 Changing trends in defect management**
The key challenge of software industry is to engineer a software product with minimum post deployment defects. Advancement in fundamental engineering aspects of software development enables I.T. enterprises to develop a more cost effective and better quality product through systematic defect detection and prevention strategies. Investing in defect prevention reduces the cost of defect detection and elimination. It is a sensible commitment towards production of quality software. Small increase in the prevention measures overall produces a major decrease in total quality cost. The main intent of quality cost analysis is not to remove the cost entirely. However, it ensures maximum benefit from the investment. The knowledge of quality cost analysis brings awareness from detection of defects to prevention of defects (Spiewak R & McRitchie K, 2008). An observation in progressive software industries prove the fact that cost to ensure quality reduces with defect detection and prevention strategies.

Changing trends in defect management enables transition from postproduction detection technique to preproduction detection technique and in situ detection during developmental phase. Here we describe some of the modern approaches in situ detection during the software development.

*Cost quality analysis through defect injection and defect removal techniques*

Recent research trend includes study of cost investment in defect injection and defect removal as a part of process maturity (Lars M. Karg & Arne Beckhaus, (2007). Phase-based defect removal model (DRM) analyses phase wise number of defects injected, number of defects removed and number of defects escaped from previous phase to the current phase.

*Defect prediction*

The industry's current interest is towards predicting the number of latent defects. Defect prediction is a technique of detecting the quality of the software before deployment. It enhances both project and product performance. The main intension is to gain complete confidence with the customers through the products. Defect prediction techniques include empirical defect prediction, defect discovery profile, Constructive Quality Model (COQUALMO), Orthogonal Defect Classification (Bard Clark & Dave Zubrow, 2001).

*Personal quality management*

Major contribution for defect occurrences is human intensive. Hence, modern strategy of defect management emphasizes upon personal quality management. It provides individual software developers as well the team to prevent and remove defects at the early stages of the development. It has a promising and positive impact on software quality. Essence of personal software process and team software process is to make quality aspect more individual responsibility and group cohesiveness (Watts S. Humphrey, 1994).

*Modern approach of testing*

Defect prevention is one of the best ways of defect management. Testing detects those defects, which has escaped the eyes of developers. It only detects presence of defects but cannot prevent them (Ian Sommerville, 2008; Srinivasan N. & P. Thambidurai, 2007; Glenford J. Myers at el. 2004). It is the slowest technique in software process for defect detection. Testing is the last opportunity to weed out the defects that is highly expensive to deal with at the later stages. Conventional classifications of testing are by purpose, by life cycle phase and by scope. Testing by purpose includes correctness testing, performance testing, reliability testing and security testing. Testing by life cycle phase includes requirements phase testing, design phase testing, implementation phase testing, evaluating test results, installation phase testing, acceptance testing and maintenance testing. Testing by scope includes unit testing, component testing, integration testing and system testing.

Modern approach of testing includes test automation. It increases quality and reduces testing cost and time. This insists a need for automation strategy to decide upon what, when and how much to automate. It requires prioritization of automation test plans too (Hung Q. Nguyen at el., 2006).

System testing using Markov chain model is an advanced testing approach. This technique of testing emphasize upon the probability of defect occurrence, probability of the usage of the functionality, most probable test for the functionality, required test coverage using postman algorithm and possibility of automating the entire testing process (Prowell, S.J., 2005).

Agile approach of software development integrates testing as a continual developmental activity. Time to market is the motto for agile approach of software development. It redefines traditional formal quality assurance activities into daily activity. Test automation is a basic requirement in this approach (Peter Schuh, 2005).

## 3. Significance of inspection technique in defect detection and prevention

With the knowledge of defect and defect management strategy, we now describe the effectiveness and efficiency of inspection technique in defect detection and prevention. It largely reduces defect migration and manifestation into later stages of development. Since three decades, inspection has proven to be the most mature, valuable and competent technique in this challenging area (Micheal Fagan, 2002; Sami Kollanus & Jussi Koskinen , 2007, Rombach at el., 2008).

### 3.1 Origin of inspection
Conventional approach of software development was not successful in delivering a defect-free product. Estimates of rework to fix a defect reported by customers ranged from 30% to 80% of total developmental effort. Hence, it was required to detect defects in the product and the process that cause defect occurrence. This led to the origin of inspections in software development. Groundwork for software inspection was in 1972 by Michael Fagan. He is the pioneer of inspections. He emphasized on inspection to be a formal activity. In his original data, he was able to detect 82% of defects during design and code inspection. By implementing software inspection in their process, he was able to save millions of dollars in developmental cost. For this reason, Fagan received largest corporate individual award (Michael Fagan, 2002).

### 3.2 Benefits of inspection
Inspection is one of the powerful techniques for the early defect detection. Inculcating the inspection activity in SDLC serves to be one of the best practises in the developmental process. Benefits reaped by implementing inspection technique in the software process are reduction of defects earlier in the product and process at less cost. It increases customer satisfaction and enhances productivity. It ships the product within the specified time, deploy high quality product and saves cost, time and developmental effort towards rework. It further reflects process maturity of software industry and builds team spirit. It also strengthens individual confidence and reduces testing time. It also serves as a defect preventive measure (Caper Jones, 2008; Roger Stewart & Lew Priven, 2008; David L. Parnas & Mark Lawford, 2003; Oliver Laitenberger, 2002; Karl E. Wiegers, 1995; Doolan E. P., 1992).

### 3.3 Inspection techniques
Michael Fagan first seeded the concept of formal inspection technique. However, many great contributions have been made in the domain of inspection. Few of these contributions are discussed as various popular inspection techniques (Bordin Sapsomboon, 1999). They are
*Fagan's Software Inspection*
Fagan's Inspection is a structured, meeting-oriented procedure, which includes the activities of overview, preparation, inspection, rework and follow-up. Inspection meeting identifies, classifies and logs all possible defects.

*Formal Technical Reviews*
Formal technical reviews consist of a group of technical personnel who cooperate with each other to analyze the artefacts of the software development process. The outcome is a structured report. The main objective is to examine the artefact, appraise it and produce a summary report. Effectiveness of reviews to reach expected quality complies with standards and guidelines in the form of checklists, forms, summary reports etc.

*Structured Walkthroughs*
Walkthrough is a peer review process carried out by a group of non managerial staff where each participant has their well specified roles like scribe, reviewer, author and so on. A structured walkthrough has a set of defined phase activities and the outcome is a list of comments or discussions made. They are a means of educating the participants in the software project.

*Code Reading*
Code reading is informal activity where a small group of participants reads the source code for defect identification at an optimal rate of approximately 1K lines per day.

*Humphrey's Inspection Model*
A specialized team carries out inspection with well-defined roles assigned to each member of the team. It is an extension of Fagan's Inspection. The model includes overview, preparation, analysis, inspection, rework and follow-up in lieu of three steps of preparation, inspection and repair. It is a structured technique of inspection.

*Formal Technical Asynchronous review method (FT Arm)*
This technique employs the parallel activities of setup, orientation, private review, public review, consolidation and group review. Result of private review is reviewer's comments for each node. In public review, each reviewer vote asynchronously for each comment through open discussion. Consolidation of the public and private review resolves the issues. Group review meeting will take up unresolved issues. Thus, all activities in FT Arm technique of inspection are asynchronous.

*Gilb Inspection*
Gilb inspection technique includes entry, planning, checking, logging, brainstorming, edit, follow-up, exit activities. Inspection process begins when certain entry criteria is met. With the identification of defects, their rectification through RCA and logging, an exit criteria is declared. This acts as a token of completion of inspection.

*Phased Inspection*
The main advantage of the phased inspection is to deliver the defect-free product by emphasising on the quality attributes like maintainability, portability, reusability etc. Phased inspection technique uses computer-supported software inspection. Each inspection phase occurs in serial fashion by either a single inspector or multiple inspectors to review. The activities involved are examination, inspection and reconciliation.

*N-fold Inspection*
In N-fold inspections, multiple inspections occur in parallel for the same artefact. The prediction is that multiple inspections can detect those defects that might have escaped from the eyes of a single inspector.

*Clean room approach*
Clean room approach is an advanced inspection technique that aims at delivering a zero or minimal defect product. The key feature of this approach is usage of mathematical reasoning for correctness proof.

### 3.4 Inspection Metrics

Metrics are numerical values that quantify the process and the product. They define, measure, manage, monitor and improve the effectiveness of the process and the product. They serve as criteria upon which the inspection planning improves (David F. Rico, 2004). The main objective of using inspection metrics is to improve on defect detection and reduce cost of rework. Identification of defect at the deployment stage or even later in the development phases is highly expensive. Cost to fix a defect found at requirement phase after deployment of the product is 100 times the cost of fixing it at the phase. Cost to fix a defect found at design phase after shipment of the product is 60x. Cost to fix a defect at implementation phase found by customers is 20x. Above cost quality analysis proves the significance of inspection in developmental process. The most commonly used inspection metrics are

$$\text{Total number of defects} = A + B - C \qquad (1)$$

Where A and B are the total number of defects detected by reviewer 1 and reviewer 2 while C is the total number of common defects detected by both reviewers.

Defect density is the ratio of the number of defects found to the size of the artefact where size can be lines of code or number of modules, number of function points etc. according to the suitability of industrial house engaged in the process.

$$\text{Defect Density} = \text{Total defects found} / \text{Size} \qquad (2)$$

Estimated Total Number of Defects is the sum of the total number of defects found and the estimated total number of defects remaining. Capture recapture approach detects total number of defects.

$$\text{Estimated Total Number of Defects} = A * B / C \qquad (3)$$
(by considering only two reviews for simplicity)

The Defect Removal Efficiency (DRE) of an inspection process is also termed as Inspection Yield, which is

$$\text{Inspection Yield} = \text{Total Defects Found} / \text{Estimated Total Defects} * 100\,\% \qquad (4)$$

Defect Removal Efficiency is a measure for the defect removal ability in the development process. This measurement can be either for the entire life cycle or for each phase of the development life cycle. Use of DRE at front end (before the code integration) is early defect removal and when used at specific phase, it is phase effectiveness (Jasmine K.S and Vasantha R. 2007).

$$\text{DRE} = (\text{Defects removed during a development phase} / \text{defects latent in the product}) * 100\% \qquad (5)$$

Latent defects is the sum of defects removed during the phase and the defects found late. Inspection Time is the total inspection time measured in hours which is given by

$$\text{Inspection Time} = \text{Sum of each reviewer's review time} + \text{total person time spent in each meeting} \tag{6}$$

Inspection Rate can be computed with the inspection time and the size of the artefact, which is measurable in terms of number of pages or LOC or other such measures as

$$\text{Inspection Rate} = \text{Size} / \text{Total Inspection Time} \tag{7}$$

The defect detection Rate is estimated based on efficiency in detecting the defects which is computed as

$$\text{Defect Finding Efficiency} = \text{Total Defects Found} / \text{Total Inspection Time} \tag{8}$$

A new metric to quantify the inspection capability is recently introduced. The metric is accepted widely and identifies itself as Depth of Inspection (DI), which is defined as

$$\text{Depth of Inspection} = \text{Total defects identified by inspection} / \text{Total defects identified by inspection and testing} \tag{9}$$

Depth of inspection yield and depth of testing yield are two new measures to quantify the efficiency of defect capturing technique.

$$\text{Depth of inspection (\%)} = \text{number of defects detected by inspection} / \text{total number of defects detected by both inspection and testing} * 100 \tag{10}$$

$$\text{Depth of testing (\%)} = \text{number of defects detected by testing} / \text{total number of defects detected by both inspection and testing} * 100 \tag{11}$$

Above equations from (1) to (11) are some of the measures that quantify the outcome of inspection activity.

## 4. Analysis of inspection process in various life cycle phases in software development

Inspection is functional at every phase of the software development to uncover maximum number of defects. This defect prevention approach has proved to be most effective and efficient among several other existing approaches.

During requirements phase, the product manager interacts with the sales person, marketing person and stakeholders to perform a comprehensive analysis and validates the product requirements. The outcome is a requirement specification also called as Product Requirement Definition (PRD). First round of inspection uncover defects found in PRD. Outcome of this inspection is a list of comments. Validated assumptions to these requirements remove ambiguity. Some of these assumptions may themselves lead to defects. A second round of inspection identifies defects due to assumptions made in requirement definition.

During design phase, the inspection artefacts are high-level design and low-level design. Inspection team thoroughly inspects the assumptions made with respect to interactions between subsystems and other such dependency factors. The outcome is identification of flaws due to lack of clarity in design.

Implementation phase begins with write up of test cases for software to be developed. Code generation prior to test case write up is always error prone. Hence, inspection of test cases ensures code generation to be defect-free.

### 4.1 Case Study

The following case study gives information on various defect detection and prevention techniques followed in different companies to deliver a high quality product. They include
   a) A leading product-based company
   b) A leading service-based company
   c) DP techniques adopted in a company that is not stringent towards defect prevention activities

**a) Effective defect prevention techniques adopted in leading product-based company**

The company follows staged process model, which is a representation of CMMI (Capability Maturity Model Integrated) Meta model. CMMI describes goals and best practises for every process area. It represents its process areas as either continuous model or staged model. Staged representation of CMMI defines five maturity levels and process areas that are required to achieve the maturity level. Each process area contains goals, common features and practices.

Since 1999-2000, the company follows qualitative and quantitative analysis as a defect preventive strategy. It maintains a database to capture all the defects including the field defect. Field defects are the mistakes identified by the customer after shipment of the product to field. Qualitative analysis comprises of stage kick off meeting prior to the start of each life cycle phase or task. Purpose of the meeting is to highlight those areas where mistakes were committed, identified and actions that were taken for their rectification in the past. Sensitization and discussions for current project is accomplished through the lessons learned from previous similar type of projects. Rationale is to educate in reducing defect injection and increasing defect removal efficiency.

Quantitative approach collects authentic and realistic data from the stored projects.

Categorization of projects follows Pareto principles of 80% rule. Accordingly, projects implemented with similar platform and technologies forms a cluster. Control chart is statistical tool that measures for consistency checks at all phases of SDLC. If an inspection effort at a phase exemplifies the non-conformance of the defects in the control band, it reveals the fact that either review was excellent or if review was reprehensible. Testing comprises of regression testing which ensures non-introduction of unintentional behaviour or additional errors in the software, performance test ascertain the performance of requirements, environmental test performs testing of operational environment of the product, health test is conducted for users of the product to verify the product in compliance with health safety standards.

The review efficiency metric gives an insight on quality of review conducted. Review efficiency is idyllic if it can identify one critical defect per every one-man hour spent on reviews.

$$\text{Review Efficiency} = \text{Total number of defects found by reviews} / \text{Total number of defects in product} \qquad (12)$$

With a review efficiency of 87%, the company has reported an increase in their productivity from 250 to 400 accentuating the importance of adopting DP strategies. With an inspection-testing time ratio of 15:30, the company was able to record a quality level of 99.75% defect-free product.

**Observation**

Table.3. depicts anatomy of inspection and testing activities for the sampled five projects. It specifies time and defect profile for the above projects. Application of recent metrics (9), (10), (11) and from the analysis on the sampled data, following observations are made.

- Inspection is functional at all phases of software development. Deliverables for inspection are requirement specification, high-level, low-level design artefacts and code reviews.
- Percentage of defect distribution at requirements phase is observed to be in the range of 50% to 60%. Defects occur in the range of 18% to 26% at design phase. Implementation phase has 12% to 19% of total defect distribution.
- Company schedules 10% to 14% of the total project time at each phase for inspections and 20% to 30% of total project time for testing to deploy defect-free product.
- With the scheduled inspection time, inspection team is able to unearth 40% to 70% of defects at the early stages of development.
- Inspection can detect only static defects. Testing is vital to detect dynamic defects. With the scheduled testing time, the company is able to detect majority of defects. It is impossible to capture all defects in any application because of varied complexities. Company claims up to 97% defect-free product.

**b) Effective defect prevention techniques adopted in leading service-based software company**

The company follows continuous representation of CMMI Meta model. It defines five capability levels and process areas to access capability levels of the company. Each process area further contains specific goals and specific practices with generic goals and generic practises.

Since 2002, the company follows defect detection and defect prevention techniques to enhance quality of the product. The defect detection techniques include review of plans, schedules and records. Company follows product and process audits as part of quality control activities to uncover defects and correct them. The defect prevention techniques followed in the company includes pro-active, reactive and retrospective DP.

Pro-active DP aims to create an environment for controlling defects rather than reacting to it. A stage kick off meeting is conducted to reveal those areas where mistakes were committed, recognized and actions that were taken for their refinement in the past. Company deems from the previous projects, the lessons learnt from the life cycle phases, the DP action items documented and best practices adopted. It emphasizes the development team to follow DP action items from the previous projects in the organization that are of same nature.

Reactive DP identifies and conducts RCA (Root Cause Analysis) for defects meeting at trigger points or at logical points. Implementation of curative actions along with preventive actions eliminates the potential defects. The most common root causes for defects identified in the company are due to lack of communication, lack of training, oversight, lack of project methodology and inappropriate planning.

Performance of retrospection towards the end of the project or at identified phases of SDLC explores areas with strong points together with areas requiring perfection.

**Observation**

Table.4. depicts anatomy of inspection and testing activities for a sampled five projects. It indicates time and defect profiles for the above projects. From the table, following observations are listed as:

- Inspection is functional at all phases of software development.
- Percentage of defect distribution at requirements phase is observed to be in the range of 48% to 61%of total defects. Defects occur in the range of 20% to 28% at design phase.
- Implementation phase contains 13% to 25% of total defect distribution.
- Company schedules 10% to 14% of the total developmental time for inspections and 21% to 35% of total time for testing at each phase to deploy defect-free product.
- Depth of Inspection yield proves that inspection team is able to unearth 37% to 69% of defects at the early stages of development.
- With the scheduled testing time, the company is able to detect majority of remaining defects.
- Company claims up to 97% defect-free product.

As a factual statement, companies adapting to DP strategies have shown that over a period of time, quality of the product enhances while the cost of quality reduces.

**c) DP techniques adopted in a company that is not stringent towards defect prevention activities**

The study also includes a company that is not strictly adhering to DP strategies in comparison with the observations made from the previous two companies.

**Observation**

Table 5. depicts anatomy of inspection and testing activities for a sampled five projects. From the table, following observations are listed as :

- Inspection is functional at all phases of software development.
- Percentage of defect distribution at requirements phase is in the range of 44% to 54%.
- Defects occur in the range of 16% to 19% at design phase. Implementation phase has 13% to 18% of total defect distribution.
- Company schedules 3% to 7% of the total project time for inspections and 50% to 55% of total time for testing at each phase to deploy defect-free product.
- Depth of Inspection yield indicates that inspection team is able to unearth only 16% to
- 33% of defects at the early stages of development.
- With the scheduled testing time, the company is able to detect majority of defects.
- Company claims up to 86% defect-free product.

Since the company is not very stringent towards DP activities, testing requires a substantial amount of time.

Figure. 1. shows a comparative graph of inspection and testing for five selected projects from three different companies. Figure.2. shows a comparative graph of inspection yield and testing yield of the three companies under study. Figure. 3. shows the defect capturing capability of the three companies with the follow of up of their DP strategies to eliminate defects. The graphs indicate efficiency of inspection in defect detection. It detects defects close to the point of injection. Further, with increase in inspection time, testing time decreases. Investment in inspection is initially high but over a period of time, the cost of quality reduces while quality increases. It further reflects the process maturity of the company.

| Company 1 | P1 | P2 | P3 | P4 | P5 |
|---|---|---|---|---|---|
| Total time( in man hours) | 250 | 300 | 500 | 4248 | 6944 |
| Requirement time | 25 | 32 | 50 | 1062 | 2597 |
| Requirement review | 3 | 4 | 5 | 107 | 281 |
| Requirement test | 7 | 9 | 15 | 320 | 621 |
| Total inspection time (%) | 12.00 | 12.5 | 10.00 | 10.08 | 10.82 |
| Total testing time (%) | 28.00 | 28.13 | 30.00 | 30.13 | 23.91 |
| Total number of defects | 30 | 46 | 70 | 175 | 254 |
| Total number of defects detected by inspection | 16 | 31 | 49 | 80 | 112 |
| Total number of defects detected by testing | 14 | 15 | 21 | 95 | 142 |
| Depth of Inspection yield (%) | 53.33 | 67.39 | 70.00 | 45.71 | 44.09 |
| Depth of Testing yield (%) | 46.67 | 32.61 | 30.00 | 54.29 | 55.91 |
| Defect percentage | 60.00 | 56.10 | 59.32 | 59.32 | 56.07 |
| Design time | 46 | 46 | 100 | 1411 | 1966 |
| Design review | 6 | 5 | 11 | 143 | 200 |
| Design test | 13 | 10 | 30 | 390 | 396 |
| Total inspection time (%) | 13.04 | 10.87 | 11.00 | 10.13 | 10.17 |
| Total testing time (%) | 28.26 | 21.74 | 30.00 | 27.64 | 20.14 |
| Total number of defects | 10 | 15 | 28 | 66 | 120 |
| Total number of defects detected by inspection | 5 | 7 | 15 | 32 | 77 |
| Total number of defects detected by testing | 5 | 8 | 13 | 34 | 43 |
| Depth of Inspection yield (%) | 50.00 | 46.67 | 53.57 | 48.48 | 64.17 |
| Depth of Testing yield (%) | 50.00 | 53.33 | 46.43 | 51.52 | 35.83 |
| Defect percentage | 20.00 | 18.29 | 23.73 | 22.37 | 26.49 |
| Implementation time | 101 | 118 | 150 | 878 | 1300 |
| Code review | 10 | 17 | 20 | 105 | 156 |
| Testing | 30 | 34 | 45 | 265 | 310 |
| Total inspection time (%) | 10 | 14 | 13 | 12 | 12 |
| Total testing time (%) | 30 | 29 | 30 | 30 | 24 |
| Total number of defects | 8 | 16 | 15 | 47 | 67 |
| Total number of defects detected by inspection | 4 | 7 | 7 | 24 | 37 |
| Total number of defects detected by testing | 4 | 9 | 8 | 23 | 30 |
| Depth of Inspection yield (%) | 50.00 | 43.75 | 46.67 | 51.06 | 55.22 |
| Depth of Testing yield (%) | 50.00 | 56.25 | 53.33 | 48.94 | 44.78 |
| Defect percentage | 16.00 | 19.51 | 12.71 | 15.93 | 14.79 |
| Total number of defects | 50 | 82 | 118 | 295 | 453 |
| Sum of defects captured | 48 | 77 | 113 | 288 | 441 |
| Total defects captured (%) | 96.00 | 93.90 | 95.76 | 97.63 | 97.35 |

Table 3. Time and defect profile of a leading product based company

| Company 2 | P1 | P2 | P3 | P4 | P5 |
|---|---|---|---|---|---|
| Total time( in man hours) | 263 | 340 | 507 | 4786 | 7416 |
| Requirement time | 26 | 40 | 55 | 2047 | 2340 |
| Requirement review | 3 | 4 | 6 | 200 | 235 |
| Requirement test | 7 | 11 | 16 | 575 | 821 |
| Total inspection time (%) | 12 | 10 | 10.91 | 9.77 | 10.04 |
| Total testing time (%) | 27 | 27.5 | 29.09 | 28.09 | 35.09 |
| Total number of defects | 35 | 50 | 77 | 200 | 420 |
| Total number of defects detected by inspection | 17 | 26 | 40 | 123 | 156 |
| Total number of defects detected by testing | 18 | 24 | 37 | 77 | 264 |
| Depth of Inspection yield (%) | 48.57 | 52.00 | 51.95 | 61.50 | 37.14 |
| Depth of Testing yield (%) | 51.43 | 48.00 | 48.05 | 38.50 | 62.86 |
| Defect percentage | 51.47 | 50.00 | 61.60 | 59.70 | 48.72 |
| Design time | 40 | 50 | 110 | 1323 | 2950 |
| Design review | 4 | 6 | 11 | 128 | 300 |
| Design test | 13 | 13 | 25 | 275 | 640 |
| Total inspection time (%) | 10 | 12 | 10 | 10 | 10 |
| Total testing time (%) | 33 | 26 | 23 | 21 | 22 |
| Total number of defects | 15 | 28 | 26 | 75 | 201 |
| Total number of defects detected by inspection | 8 | 12 | 14 | 33 | 86 |
| Total number of defects detected by testing | 7 | 16 | 12 | 42 | 115 |
| Depth of Inspection yield (%) | 53.33 | 42.86 | 53.85 | 44.00 | 42.79 |
| Depth of Testing yield (%) | 46.67 | 57.14 | 46.15 | 56.00 | 57.21 |
| Defect percentage | 22.06 | 28.00 | 20.80 | 22.39 | 23.32 |
| Implementation time | 100 | 130 | 165 | 756 | 956 |
| Code review | 10 | 14 | 23 | 91 | 116 |
| Testing | 35 | 30 | 45 | 165 | 235 |
| Total inspection time (%) | 10 | 10.77 | 13.94 | 12.00 | 12.13 |
| Total testing time (%) | 35 | 23.08 | 27.27 | 21.83 | 24.58 |
| Total number of defects | 14 | 20 | 17 | 53 | 219 |
| Total number of defects detected by inspection | 8 | 8 | 9 | 27 | 152 |
| Total number of defects detected by testing | 6 | 12 | 8 | 26 | 67 |
| Depth of Inspection yield (%) | 57.14 | 40.00 | 52.94 | 50.94 | 69.41 |
| Depth of Testing yield (%) | 42.86 | 60.00 | 47.06 | 49.06 | 30.59 |
| Defect percentage | 20.59 | 20.00 | 13.60 | 15.82 | 25.41 |
| Total number of defects | 68 | 100 | 125 | 335 | 862 |
| Sum of defects captured | 64 | 98 | 120 | 328 | 840 |
| Total defects captured (%) | 94 | 98 | 96 | 98 | 97 |

Table 4. Time and defect profile of a leading service based company

| Company3 | P1 | P2 | P3 | P4 | P5 |
|---|---|---|---|---|---|
| Total time( in man hours) | 150 | 225 | 368 | 490 | 550 |
| Requirement time | 15 | 20 | 30 | 54 | 55 |
| Requirement review | 1 | 1 | 2 | 3 | 3 |
| Requirement test | 8 | 10 | 16 | 30 | 30 |
| Total inspection time (%) | 6.67 | 5.00 | 6.67 | 5.56 | 5.45 |
| Total testing time (%) | 53.33 | 50.00 | 53.33 | 55.56 | 54.55 |
| Total number of defects | 15 | 25 | 60 | 65 | 80 |
| Total number of defects detected by inspection | 4 | 4 | 14 | 16 | 19 |
| Total number of defects detected by testing | 11 | 21 | 46 | 49 | 61 |
| Depth of Inspection yield (%) | 26.67 | 16 | 23.33 | 24.62 | 23.75 |
| Depth of Testing yield (%) | 73.33 | 84 | 76.67 | 75.38 | 76.25 |
| Defect percentage | 50 | 54.35 | 48.39 | 50.00 | 44.44 |
| Design time | 30 | 35 | 42 | 70 | 77 |
| Design review | 1 | 2 | 2 | 4 | 5 |
| Design test | 15 | 18 | 22 | 36 | 40 |
| Total inspection time (%) | 3.33 | 5.71 | 4.76 | 5.71 | 6.49 |
| Total testing time (%) | 50 | 51.43 | 52.38 | 51.43 | 51.95 |
| Total number of defects | 5 | 8 | 20 | 24 | 35 |
| Total number of defects detected by inspection | 1 | 2 | 6 | 6 | 10 |
| Total number of defects detected by testing | 4 | 6 | 14 | 18 | 25 |
| Depth of Inspection yield (%) | 20.00 | 25.00 | 30.00 | 25.00 | 28.57 |
| Depth of Testing yield (%) | 80.00 | 75.00 | 70.00 | 75.00 | 71.43 |
| Defect percentage | 16.67 | 17.39 | 16.13 | 18.46 | 19.44 |
| Implementation time | 45 | 85 | 105 | 180 | 165 |
| Code review | 2 | 4 | 6 | 10 | 9 |
| Testing | 24 | 45 | 57 | 94 | 85 |
| Total inspection time (%) | 4.44 | 4.71 | 5.71 | 5.56 | 5.45 |
| Total testing time (%) | 53.33 | 52.94 | 54.29 | 52.22 | 51.52 |
| Total number of defects | 4 | 6 | 21 | 23 | 27 |
| Total number of defects detected by inspection | 1 | 2 | 7 | 6 | 8 |
| Total number of defects detected by testing | 3 | 4 | 14 | 17 | 19 |
| Depth of Inspection yield (%) | 25 | 33.33 | 33.33 | 26.09 | 29.63 |
| Depth of Testing yield (%) | 75 | 66.67 | 66.67 | 73.91 | 70.37 |
| Defect percentage | 13.33 | 13.04 | 16.94 | 17.69 | 15.00 |
| Total number of defects | 30 | 46 | 124 | 130 | 180 |
| Sum of defects captured | 24 | 39 | 101 | 112 | 142 |
| Total defects captured (%) | 80.00 | 84.78 | 81.45 | 86.15 | 78.89 |

Table 5. Time and defect profile of a company not stringent to DP

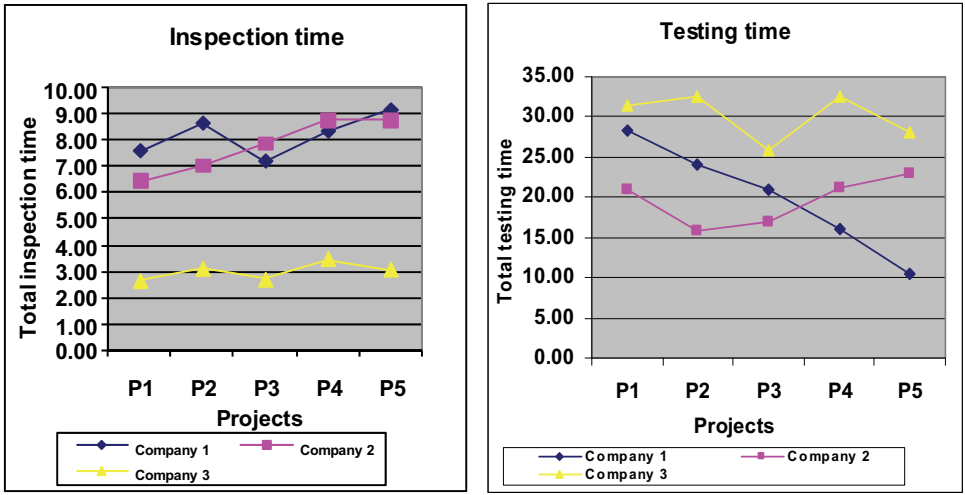

Fig. 1. Comparative graphs of inspection and testing for three companies over five selected projects

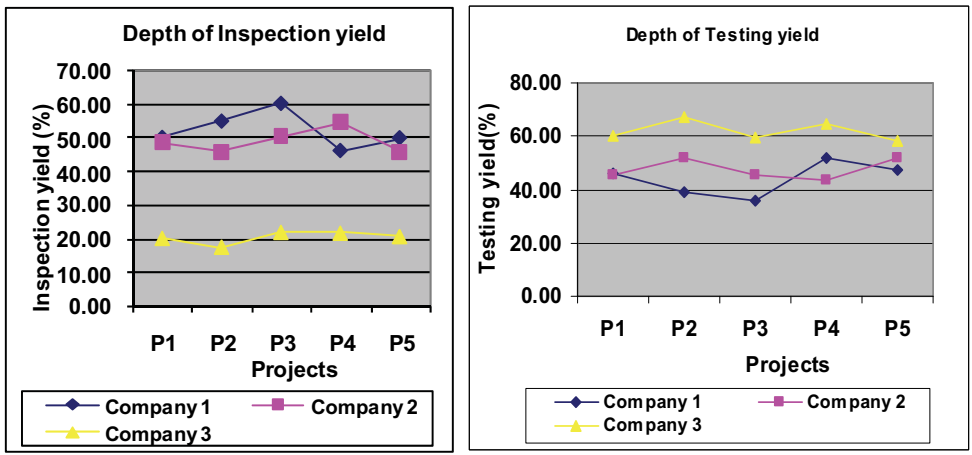

Fig. 2. Comparative graphs of inspection yield and testing yield for three companies over five selected projects

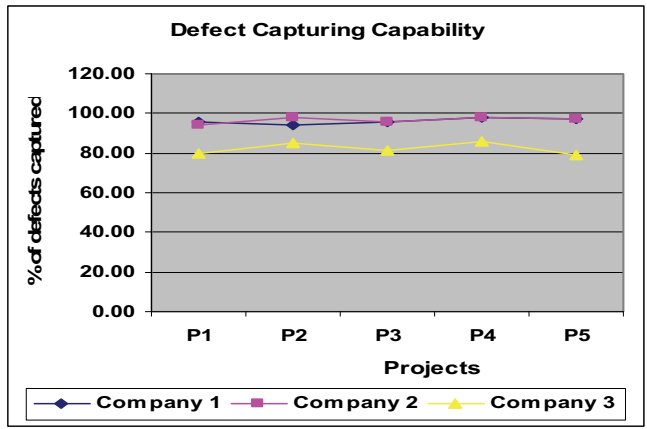

Fig. 3. Defect capturing capability of the three companies

## 5. Parameters influencing inspection

An empirical study of various projects across several service-based and product-based companies has revealed the impact of certain parameters on effectiveness of inspection activity at every phase of SDLC. These parameters are inspection time, number of inspectors involved, experience of inspectors involved at each phase of software development and preparation (training) time.

Inspection time is a highly influencing parameter. Framework of time for inspection at all phases of development life cycle is necessary. The above Case Study proves importance of inspection and emphasize upon scheduling appropriate inspection time. Development of 99% defect-free product is possible with an inspection time of 10%-15% out of total project time (Suma V & Gopalkrishnan Nair T.R., 2008a; Suma V & Gopalakrishnan Nair T.R., 2008b). Reduction in inspection time can cause defects to escape from identification. However, automation of software inspection can reduce manual inspection time and retain the effectiveness of defect detection (Jasper Kamperman, 2004).

Number of inspectors influences inspection process at each phase of development (Halling M & Biffl S, 2002; Stefen Biffl & Michael Halling, 2003 b). Self-reviewer in compliance with self-review checklist will initially inspect each deliverable. Peer review detects defects that have escaped from the eyes of the author. Hence, peer review is an effective defect detection and removal activity (Karl E.Weigers, 2001b; Karl E. Weigers, 2002; Steven H Lett, 2007). It can be either formal or informal activity. Outcome of inspection is inspection report. It is a list of comments with identified defects. Concerned authors receive the inspection report for fixing up of defects. Thus, feedback mechanism facilitates developer team and management to identify and remove defects along with fault processes. Identified defects further undergo causal analysis before final inspection. The technical leader performs root cause analysis of defects. Inspection team maintains a record of defects that includes type of defect, number of defects, root causes of defects, inspectors involved, their experience level as inspectors and action items taken for rectification of the identified defects and so on. This log acts as a lesson learnt for future development of the same project or for projects that are similar in nature. Project leaders and technical managers or senior managers perform the final inspection. Project leaders responsible for the deliverable should not be the inspector for the final inspection. Typically, some organizations prefer two inspectors namely self-inspector and peer inspector. However, effectiveness of inspection can occur only when team size is proportional to the size and complexity of the project. Effectiveness of inspection can be further accelerated with division of responsibilities. Selection of team size also depends upon company's budgetary strategies.

Experience of inspectors is an influencing parameter in defect detection. Established projects require lesser time in elicitation of requirement than innovative projects. Hence, innovative projects demand experienced inspectors. Integrated projects need more developmental time in design phase than other life cycle phases. Such projects demand experienced inspectors at design phase. An inspector who has examined design deliverables for a minimum of three projects is preferred for inspecting the high-level design and low-level design. Inspection of design artefacts includes examining the conformance to security, maintainability, reusability, complexity of design and other such related issues. An inspector at implementation phase requires knowledge on security aspects and intuitive knowledge on rules of the organization. He needs coding experience along with ability to check for redundancy in code, number of lines of code, code efficiency, design policies against code, security, safety, maintainability, reusability and other such quality attributes. Besides, human factors influence effectiveness of inspection. Organizational and self-motivational factors have impact on defect detection. The competence of inspectors turns out to be a very important component for effective inspection process.

Preparation (Training) time for inspectors plays a vital role in influencing the effectiveness of inspection. Preparation time varies depending on the complexity of the project. Author conducts a walkthrough of the deliverable to the participating inspectors. This helps team members for analysing defects. Improvement in competency of inspectors and reduction in human effort can be accomplished by the inspection team through formal training, education or through perspective-based reading techniques. They specify what to and how to inspect the artefacts (Biffl S. 2000; Stefan Biffl at el., 2003a; Stefen Biffl & Michael Halling, 2003 b; Karl E. Wiegers, 2001a). Thus, inspection time and preparation time influences efficiency of inspection meeting (Liguo Yu at el., 2006). However, our empirical study reveals that inspection time, number of inspectors and their experience with preparation time influences defect detection rate.

Models such as Baysian belief network builds confidence in measuring the effectiveness of inspection. The strength of this model is in the usage of attributes such as product size, complexity of product, quality of inspection process that contributes towards inspection effectiveness (Trevor Cockram, 2001). Code inspection model uses code size, number of coding errors and density of coding errors to estimate the effectiveness of inspection.

## 6. Conclusions

Software has strong influence on all categories of occupations. The key challenge of IT industry is to deploy high quality defect-free software product. Software Engineering foundation helps engineers to develop defect-free software within the scheduled time, cost and resources in a systematic manner. Defect is an undesirable behaviour or an imperfection that occurs in the product through the process. Hence, defect management is the core business need of the day. Two successful approaches of defect management are defect detection and defect prevention. Empirical studies reveal the probability of percentage of defect distribution at every phase of software development. Study specifies that 50% to 60% of total defects originate at requirement phase, 15% to 30% at design phase and 10% to 20% of total defects occur at implementation phase. Analysis of defect pattern indicates existence of various categories of defects and possibilities of their occurrences. Blocker defect occur in the range of 5% to 15%, major defect in 10% to 25%, critical defect in 20% to 25%, minor defect in 10% to 20% and trivial defect in 15% to 55% out of total percentage of defects.

Defect detection is a technique of identifying defects and eliminating them at the root level. Some of the conventional defect detection techniques are inspection, prototype, testing and correctness proof. However, an awareness of rationale for defect distribution and root cause analysis of defect pattern helps in effective defect detection. Common root causes for defect injection are analyzed to be due to lack of communication (20% to 25%), lack of education (20% to 25%), oversight (30% to 45%), transcription (20% to 25%), miscellaneous (5% to 10%).

Defect prevention is a step towards process maturity. Knowledge of defect injection methods enables defect prevention. It prevents reoccurrences, propagation and manifestation of defects either in the same project or in similar projects. Conventional defect prevention strategies include product approach, process approach and automation of developmental process. Changing trends in defect management enables transition from postproduction detection technique to preproduction detection technique and in-situ detection during developmental phase.

Inspection continues to prove as the most effective and efficient technique of defect detection and prevention since three decades. Michael Fagan is the pioneer of formal inspection technique in software industry. However, several inspection techniques have evolved over the time. They include Humphrey's inspection, Gilb inspection, formal technical reviews, structured walkthrough etc. Inspection related metrics statistically measures the effectiveness of inspections. They indicate the level of maturity of the company.

A Case Study comprising of three companies throws light on the possibility of software industries to position itself through aptly organized defect detection and prevention strategies. It proves the significance of inspection at every phase of software development. Inspection discovers static defects close to its origin while testing detects dynamic defects. Effective inspection demands 10% to 15% of total developmental time as against the testing time of 30% to 35% of total project time to unearth defects. Observation strongly indicates that software industry can deliver up to 99% defect-free products from above specified inspection and testing ratio. Reduction in inspection time demands more than 50% of total developmental time towards testing. Hence, organizations should go for defect detection and prevention strategies for a long-term Return on Investment (ROI).

An insight of parameters influencing effectiveness of inspection is a necessity to gain complete benefit of inspection technique. Appropriate inspection time, number of inspectors, experience level of inspectors and preparation time for inspectors plays a significant role in enhancing effectiveness of inspection.

The goal of reaching a consistently 99% defect-free software depends much on effective defect detection and prevention techniques adopted in the organization.

## 7. References

Anthony Mark Doggett, (2004). Statistical Comparison of Three Root Cause Analysis Tools, *Journal of Industrial Technology*, Volume 20, Number 2, February 2004 to April 2004

Bard Clark; Dave Zubrow, (2001). How Good Is the Software: A Review of Software Prediction Techniques , Software Engineering Symposium 2001, pp. 33 , sponsored by U.S. Department of Defence, Carnegie Mellon University

Biffl S., (2000). Analysis of the Impact of Reading Technique and Inspector Capability on Individual Inspection Performance, *proceedings of Seventh Asia-Pacific Software Engineering Conference APSEC 2000*, ISBN 0-7695-0915-0, December 2000, IEEE Computer Society publisher, Singapore

Bordin Sapsomboon, (1999). Software Inspection and Computer Support, state of the art paper, 1999. www.sis.pitt.edu/~cascade/bordin/soa_inspection.pdf

Caper Jones, (2008). Measuring Defect Potentials and Defect Removal Efficiency, *CROSSTALK, The journal of Defence Software Engineering*, June 2008 issue, Vol 21, No 6

Chillarege; I.S. Bhandari; J.K. Chaar; M.J. Halliday; D.S. Moebus; B.K. Ray & M.-Y. Wong, (1992). Orthogonal Defect Classification-A Concept for In-Process Measurements, *IEEE Transactions on Software Engineering*, vol. 18, no. 11, . 943-956, November, 1992

David F. Rico (Forward by Roger S. Pressman), (2004). *ROI of Software Process Improvement Metrics for Project Managers and Software Engineers,* J. Ross Publishing, ISBN: 1-932159-24-X, January 2004, Chapter 7 - Software Inspection Process ROI Methodology, USA

David N. Card, (2006). Myths and Strategies of Defect Causal Analysis, *Proceedings of Twenty-Fourth Annual Pacific Northwest Software Quality Conference*, October 10-11, 2006, Portland, Oregon, pp. 469-474, published in IT Metrics and Productivity Strategies, August 21, 2007

David L. Parnas & Mark Lawford, (2003). Guest Editors' Introduction: Inspection's role in software quality assurance, *IEEE Software Journal*, IEEE Computer Society publisher, Los Alamitos, CA, USA ,vol 20, Issue 4, pp. 16-20, 2003, ISSN:0740-7459

Doolan E. P. (1992). Experience with Fagan's Inspection Method, *Software Practice And Experience (SPE) Journal,* Wiley Publishing, Vol. 22(2), pp. 173–182, February 1992

Elfriede Dustin; Jeff Rashka & John Paul, (1999). *Automated Software Testing: introduction, management, and performance,* Addison-Wesley publisher, ISBN 0201432870, 9780201432879

Elfriede Dustin; Thom Garrett & Bernie Gauf, (2009). *Implementing Automated Software Testing: How to Save Time and Lower Costs While Raising Quality*, March 04, 2009, Addison-Wesley Professional publisher, Print ISBN- 10: 0-321-58051-6, Print ISBN-13:978-0-321-58051-1, Web ISBN-10: 0-321-61960-9, Web ISBN-13: 978-0-321-61960-0

Glenford J. Myers; Tom Badgett; Corey Sandler & Todd M. Thomas, (2004). *The Art of Software Testing*, Second Edition, 2004, John Wiley & Sons publisher, ISBN 047167835X, 9780471678359

Halling M & Biffl S, (2002). Investigating the influence of software inspection process parameters on inspection meeting performance, *Proceedings of International Conference on Empirical Assessment of Software Engineering (EASE), Keele,* Volume 149, Issue 5, Oct 2002 pp. 115 – 121, ISSN: 1462-5970, The Institution of Engineering and Technology publisher, United Kingdom

Hung Q. Nguyen; Michael Hackett & Brent K. Whitlock, (2006). Hay About Global Software Test Automation: A Discussion of Software Testing for Executives, August 1, 2006, ISBN-10: 1600050115, ISBN-13: 978-1600050114

Ian Somerville, (2006). *Software Engineering*, 8th Edition, ISBN-10: 0321313798, ISBN-13: 978-0321313799, Addison Wesley publisher, June 4, 2006

Jasper Kamperman,(2004). Automated Software Inspection: A New Approach to Increased Software Quality and Productivity, *Technical-White Paper*, 2004-2006

Jasmine K.S & Vasantha R., (2007) . DRE – A Quality Metric for Component based Software Products, *proceedings of World Academy Of Science, Engineering And Technonolgy*, Vol 23, ISSN 1307-6884, August 2007, Berlin, Germany

Jeff Tian, (2001). Quality Assurance Alternatives and Techniques: A Defect-Based Survey and Analysis, *Software Quality Professional*, Vol 3, No 3/2001, ASQ by Department of Computer Science and Engineering, Southern Methodist University

Jeff Tian, (2005). *Software Quality Engineering: Testing, Quality Assurance, and Quantifiable Improvement*, February 2005, Wiley John & Sons publisher, ISBN-13: 9780471713456

Karl E. Wiegers, (1995). Improving quality though software inspections, *Software Development magazine*, vol. 3, no. 4, April 1995, ISBN 716-377-5110 Visit: www.processimpact.com

Karl E. Wiegers, (2001a). Improving software inspections by using reading techniques, *Proceedings of the 23rd International Conference on Software Engineering*, Canada Pages: 726 - 727 , Year of Publication: 2001, ISBN ~ ISSN:0270-5257 , 0-7695-1050-7, Toronto, Ontario

Karl E.Weigers, (2001b). When Two Eyes Aren't Enough, *Software Development magazine*, vol. 9, no.10, October 2001

Karl E. Weigers, (2002). *Peer Reviews in Software: A Practical Guide*, Addison- Wesley publication, ISBN 0-201-73485-0, 2002

Kashif Adeel; Shams Ahmad & Sohaib Akhtar, (2005). Defect Prevention Techniques and its Usage in Requirements Gathering-Industry Practices, *Proceedings of Engineering Sciences and Technology, SCONEST*, ISBN 978-0-7803-9442-1, pp.1-5,August 2005, IEEE Computer Society publisher

Lars M. Karg & Arne Beckhaus, (2007). Modeling Software Quality Costs by Adapting Established Methodologies of Mature Industries, *Proceedings of IEEE International Conference on Industrial Engineering and Engineering Management (IEEM)*, ISBN 078-1-4244-1529-8, pp. 267-271, 2-5 December 2007, IEEE Computer Society publisher, Singapore

Liguo Yu; Robert P. Batzinger & Srini Ramaswamy, (2006). A Comparison of the Efficiencies of Code Inspections in Software Development and Maintenance, *Proceedings of The 2006 World Congress in computer Science, Computer Engineering and Allied computing, World Academy of Science Software Engineering Research and Practice*, June 26th to 29th, 2006, Las Vegas, Nevada, USA

Micheal Fagan, (2002). Reviews and Inspections, *Proceedings of Software Pioneers and Their Contributions to Software Engineering sd & m Conference on Software Pioneers*, pp. 562-573, book title Software Pioneers, ISBN-10: 3540430814, ISBN-13: 978-3540430810, 2002, Springer publisher

Mukesh Soni,(2006). Defect Prevention: Reducing Costs and Enhancing Quality, iSixSigma.com Publisher, Publishing date 19 July 2006, http://software.isixsigma.com/library/content/c060719b.asp

Oliver Laitenberger, (2002). A Survey of Software Inspection Technologies, *Handbook on Software Engineering and Knowledge Engineering,* Volume 2, 33 articles, ISBN: 981-02-4974-8, ISBN: 981-02-4514-9, 2002

Pankaj Jalote, (2002). *Software Project Management in Practice*, Addison-Wesley Professional Publishers, Jan 31, 2002, First edition, ISBN-10: 0-201-73721-3, ISBN-13: 978-0-201-73721-9

Peter Schuh, (2005). *Integrating agile development in the real world*, ISBN-10: 1584503645, ISBN-13: 9781584503644, Cengage Learning publisher, 2005

Prowell S. J., (2005). Using Markov Chain Usage Models to Test Complex Systems, *System Sciences, 2005, HICSS apos; 05, Proceedings of 38th Annual Hawaii International Conference on System Sciences*, pp. 318c - 318c, ISSN: 1530-1605, ISBN: 0-7695-2268-8, IEEE Computer Society publisher, Washington, DC, USA, 03-06 Jan. 2005, Big Island, HI, USA

Purushotham Narayan, (2003). Software Defect Prevention in a Nut shell, iSixSigma.com publisher, Publishing date 11 June 2003, http://software.isixsigma.com/library/content/c030611a.asp

Rombach; Dieter & Ciolkowski; Marcus & Jeffery; Ross & Laitenberger: Oliver & McGarry; Frank & Shull; Forrest, (2008). Impact of research on practice in the field of inspections, reviews and walkthroughs: learning from successful industrial uses, *SIGSOFT Software Engineering Notes,* Vol 33, no. 6, 2008, ISSN 0163-5948, pp. 26-35, ACM publishers, New York, NY, USA

Roger S. Pressman, (2005). Software Engineering: A Practitioner's Approach, Sixth edition, Mc Graw Hill publisher, ISBN 007-124083-7,

Roger Stewart & Lew Priven, (2008). How to Avoid Software Inspection Failure and Achieve Ongoing Benefits, *Crosstalk, The journal of Defence Software Engineering*, Volume 2, No.1, 76th Software Maintenance Group, Oklahoma Air Logistics Centre, Jan 2008

Sami Kollanus & Jussi Koskinen, (2007). A Survey of Software Inspection Technologies Research: 1991- 2005, *Working Papers* WP-40, ISBN 978-951-39-2776-9 , ISSN 0359-8489, pp. 39, March 2007, Department of Computer Science and Information Systems University of Jyväskylä, Jyväskylä, Finland

Srinivasan N. & Thambidurai P., (2007). On the Problems and Solutions of Static Analysis for Software Testing, *Asian Journal of Information Technology*, 6(2): 258- 262, Medwell Journals publisher, 2007

Spiewak R & McRitchie K, (2008). Using Software Quality Methods to Reduce Cost and Prevent Defects, *CROSSTALK, The journal of Defense Software Engineering*, Vol 21, No. 12, Dec 2008 issue

Stefan Biffl; Michael Halling & Sabine Koszegi, (2003 a). Investigating the Accuracy of Defect Estimation Models for Individuals and Teams Based on Inspection Data, *Proceedings of the 2003 International Symposium on Empirical Software Engineering (ISESE'03)*, ISESE book title, pp.232-243, ISBN 0-7695-2002-2/03, IEEE Computer Society publisher, 30 September - 1 October 2003, Rome, Italy

Stefen Biffl & Michael Halling, (2003 b). Investigating the Defect Detection Effectiveness and Cost Benefit of Nominal Inspection Teams, *Proceedings of 2003 International Symposium on Empirical Software Engineering (ISESE 2003)*, ISESE book title, Vol 29, No.5, pp.385-397, ISBN 0-7695-2002-2, IEEE Computer Society publisher, IEEE Transactions on Software Engineering, 30 September - 1 October 2003, Rome, Italy

Steven H Lett, (2007). Using Peer Review Data to Manage Software Defects, *IT Metrics and Productivity Journal*, August 21, 2007

Stefan Wagner, (2008). Defect Classification and Defect Types Revisited, Proceedings of International Symposium on Software Testing and Analysis workshop on Defects in large software systems, pp.39-40, ISBN: 978-1-60558-051-7, July 20, 2008, Washington, USA

Suma V & Gopalakrishnan Nair T. R., (2008a). Effective Defect Prevention Approach in Software Process for Achieving Better Quality Levels, *Proceedings of Fifth International Conference on Software Engineering,* pp. 2070-3740, Vol 32, ISSN 2070-3740, 30-31st August 2008, World Academy of Science, Engineering & Technology, Singapore

Suma V & Gopalakrishnan Nair T.R., (2008b). Enhanced Approaches in Defect Detection and Prevention Strategies in Small and Medium Scale Industries, *Proceedings of The Third International Conference on Software Engineering Advances*, ICSEA, pp.389 – 393, 26-31st October 2008, IEEE Computer Society publisher, Malta, Europe

Trevor Cockram, (2001). Gaining confidence with Beysian network, *Software Quality Journal*, ISSN 0963-9314 (Print) 1573-1367,(Online) ,Volume 9, Number 1/January, 2001, Springer Publishers, Netherlands

Van Moll, J.H.; Jacobs, J.C.; Freimut, B.; Trienekens, J.J.M., (2002). The importance of life Cycle modelling to defect detection and prevention, *Proceedings of 10th International Workshop on Software Technology and Engineering Practice, STEP,* pp. 144 – 155, ISBN: 0-7695-1878-8, 6-8 October 2002, Montréal, Canada

Vasudevan S, (2005). Defect Prevention Techniques, and Practices, *Proceedings of Fifth Annual International Software Testing Conference*, 2005, Hyderabad, India Visit ready test go

Watts S. Humphrey, (1989). *Managing the Software Process, Defect Prevention*, ISBN 0201180952, 9780201180954, Addison-Wesley Professional publisher, 1989

Watts S. Humphrey, (1994). A Personal Commitment to Software Quality, *Ed Yourdon's American Programmer journal*, Issue December 1994

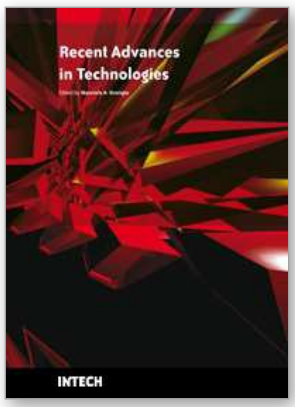

**Recent Advances in Technologies**
Edited by Maurizio A Strangio

ISBN 978-953-307-017-9
Hard cover, 636 pages
**Publisher** InTech
**Published online** 01, November, 2009
**Published in print edition** November, 2009

The techniques of computer modelling and simulation are increasingly important in many fields of science since they allow quantitative examination and evaluation of the most complex hypothesis. Furthermore, by taking advantage of the enormous amount of computational resources available on modern computers scientists are able to suggest scenarios and results that are more significant than ever. This book brings together recent work describing novel and advanced modelling and analysis techniques applied to many different research areas.

**How to reference**
In order to correctly reference this scholarly work, feel free to copy and paste the following:

Suma V and Gopalakrishnan Nair T.R. (2009). Defect Management Strategies in Software Development, Recent Advances in Technologies, Maurizio A Strangio (Ed.), ISBN: 978-953-307-017-9, InTech, Available from: http://www.intechopen.com/books/recent-advances-in-technologies/defect-management-strategies-in-software-development

# INTECH
open science | open minds

**InTech Europe**
University Campus STeP Ri
Slavka Krautzeka 83/A
51000 Rijeka, Croatia
Phone: +385 (51) 770 447
Fax: +385 (51) 686 166
www.intechopen.com

**InTech China**
Unit 405, Office Block, Hotel Equatorial Shanghai
No.65, Yan An Road (West), Shanghai, 200040, China
中国上海市延安西路65号上海国际贵都大饭店办公楼405单元
Phone: +86-21-62489820
Fax: +86-21-62489821